\documentclass[prl,aps,twocolumn,amsmath,amssymb,showpacs]{revtex4}
\usepackage{graphicx}
\begin{document}
\title{Close-packed floating clusters: granular hydrodynamics beyond the freezing point?}
\author{Baruch Meerson$^{1}$, Thorsten P\"{o}schel$^{2}$ and Yaron Bromberg$^{3}$}
\affiliation{$^1$Racah Institute of Physics, Hebrew University of
Jerusalem, Jerusalem 91904, Israel}
\affiliation{$^{2}$Institut
f\"ur Biochemie, Charit\'e, Monbijoustr. 2, 10117 Berlin, Germany}
\affiliation{$^3$School of Physics and Astronomy, Tel Aviv
University, Tel Aviv 69978, Israel}
\begin{abstract}
Monodisperse granular flows often develop regions with hexagonal
close packing of particles. We investigate this effect in a system
of inelastic hard spheres driven from below by a ``thermal" plate.
Molecular dynamics simulations show, in a wide range of
parameters, a close-packed cluster supported by a low-density
region. Surprisingly, the steady-state density profile, including
the close-packed cluster part, is well described by a variant of
Navier-Stokes granular hydrodynamics (NSGH). We suggest a simple
explanation for the success of NSGH beyond the freezing point.
\end{abstract}
\maketitle

Continuum modeling of flow of macroscopic grains remains a
challenge \cite{Kadanoff,Mills,Aranson1,Bocquet,Aranson2}. The
best known version of continuum theory here is the Navier-Stokes
granular hydrodynamics (NSGH) for a system of inelastic hard
spheres \cite{Haff, Jenkins}. The applicability of NSGH is limited
to \textit{rapid} granular flows \cite{Campbell}. By definition,
these flows are dominated by binary particle collisions, while
multi-particle interactions are negligible. Despite this drastic
simplification, the validity of the NSGH demands several
additional assumptions, some of which can be rather stringent.
Under the {\em molecular chaos} assumption, the NSGH is derivable
systematically from more fundamental kinetic equations for
inelastic hard spheres \cite{Jenkins,Sela,Brey1}. Going over from
kinetic equations to hydrodynamics, one should assume scale
separation: the mean free path of the particles should be much
smaller than the characteristic length scale, and the mean time
between two consecutive collisions much shorter than any
characteristic time scale, described hydrodynamically. The
inelasticity of particle collisions brings immediate
complications. Already at moderate inelasticity $q=(1-r)/2$ (where
$r$ is the coefficient of normal restitution of the particle
collisions), the scale separation may break down, even in the
low-density limit \cite{Grossman,Goldhirsch}. The normal stress
difference \cite{Goldhirsch} and deviations of the particle
velocity distribution \cite{Grossman,Poeschel} from the Maxwell
distribution also become important for moderately inelastic
collisions. Therefore, the NSGH is expected to be accurate only
for small inelasticity, $q \ll 1$.

Additional complications appear at large densities. Here the {\em
molecular chaos} assumption breaks down, already for
\textit{elastic} hard spheres, when the packing fraction
approaches the freezing point value $\phi_f \simeq 0.49$ (in three
dimensions) or $0.69$ (in two dimensions).  As the kinetic
equations become invalid, the constitutive relations (CRs),
necessary for the closure of hydrodynamics, are not derivable from
first principles anymore. This is the regime considered in this
work. We consider an ensemble of monodisperse, nearly elastic hard
spheres in such conditions that the standard NSGH
\cite{Haff,Jenkins} breaks down because of large densities,
\textit{not} large inelasticity. Our main objective is to check
whether a variant of NSGH can still be used in an extreme case
when the packing fraction is close to the maximum possible value,
corresponding to hexagonal packing of spheres.

We will focus on granular materials fluidized by a rapidly
vibrating bottom plate in a gravity field. Vibrofluidized granular
materials exhibit fascinating pattern-formation phenomena that
have attracted much recent interest \cite{patterngran}. In the
high-frequency and small-amplitude limit of vibrofluidization,
there is no direct coupling between the vibration and the
collective granular motion. In a simplified description of this
limit one specifies a constant granular temperature at an immobile
bottom plate. In a wide range of parameters, molecular dynamics
(MD) simulations of this system show an (almost) close-packed
cluster of particles, floating on a low-density fluid, see below.
The close-packed floating cluster is an extreme form of the
\textit{density inversion}, a phenomenon well known in
vibrofluidized granular materials. Lan and Rosato
\cite{LanRosato95} were apparently the first to observe density
inversion in three-dimensional MD simulations. Kudrolli \textit{et
al}. \cite{Kudrolli} observed a floating cluster in a
reduced-gravity experiment: a slightly tilted two dimensional
system of steel spheres rolling on a smooth surface and driven by
a vibrating side wall. Recently, a pronounced density inversion
has been observed, in two- and three dimensional vibrofluidized
granular beds, by Wildman \textit{et al}. \cite{Wildman2001}.

\begin{figure}[ht]
  \begin{center}
    \begin{minipage}[b]{4.5cm}
      \includegraphics[height=5cm, clip=]{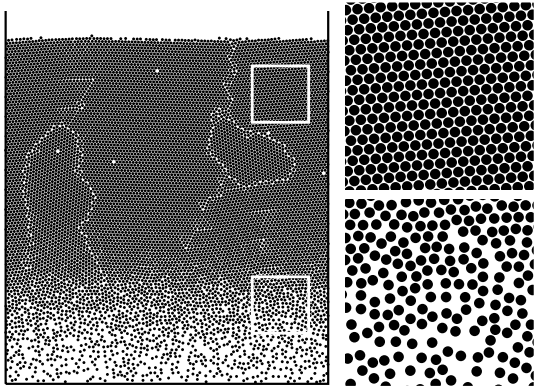}
    \end{minipage}
    \begin{minipage}[b]{2.4cm}
      \includegraphics[height=2.5cm, bb= 343 364 414 435, clip=]{1left.eps}\\[0.1cm]
      \includegraphics[height=2.5cm, bb= 343 075 414 146, clip=]{1left.eps}
  \end{minipage}
  \end{center}
\caption{A snapshot showing the close-packed floating cluster. The
parameters are $N=10^4$, $L_x=100$, $T_0=2\sqrt{3}\cdot 10^2$,
$r=0.98815$ and $g=d=m=1$. The figures on the right are
magnifications of the indicated areas.} \label{snapshot}
\end{figure}

An accurate hydrodynamic description of almost close-packed
floating clusters seems a very difficult task, as the packing
fraction here is far beyond the freezing point. Still, we will
attempt to use a variant of NSGH for this purpose. This attempt
will prove successful, and we will suggest an explanation. Here is
the model problem we are working with. Let $N \gg 1$ nearly
elastic hard spheres of diameter $d$ and mass $m$ move in a
two-dimensional box with periodic boundary conditions in
$x$-direction (period $L_x$) and infinite height. The driving base
is located at $y=0$. Gravity acceleration $g$ acts in negative $y$
direction. Upon collision with the base, the particle velocity is
drawn from a Maxwell distribution with temperature $T_0$ (which is
measured in the units of energy). The kinetic energy of the
particles is being lost by inelastic hard-core collisions
parameterized by a constant inelasticity parameter $q \ll 1$.
Figure \ref{snapshot} shows a typical snapshot of an almost
close-packed floating cluster, observed in an event-driven MD
simulation of this system. Hexagonal packing is apparent in Fig.
\ref{snapshot} \cite{defects}.

Going over to a hydrodynamic description of zero-mean-flow steady
states, we introduce coarse-grained fields: the particle number
density $n (y)$, the granular temperature $T(y)$ and the pressure
$p(y)$. The maximum possible value of $n$ is the hexagonal
close-packing value $n_c = 2/(\sqrt{3} d^2)$. A laterally uniform
steady state is described by the momentum and energy balance
equations:
\begin{equation}
\frac{dp}{dy}  + m\, n \,g = 0\,,\,\,\,\,\,\, \frac{d}{dy}
\left(\kappa \, \frac{d}{dy} T\right) - I = 0\,, \label{govern}
\end{equation}
where $\kappa$ is the thermal conductivity and $I$ is the energy
loss rate by collisions. To proceed, we need CRs: an equation of
state (EOS) $p=p\,(n,T)$ and relations for $\kappa$ and $I$ in
terms of $n$ and $T$. First-principles CRs are available only in
the low-density limit (well below the freezing point). Grossman
\textit{et al.} \cite{Grossman} derived a set of approximate {\em
global} CRs for a version of NSGH that assumes nearly elastic
collisions, but is not limited to low densities. Grossman
\textit{et al}. employed free volume arguments in the close
vicinity of the hexagonal packing and suggested simple
interpolations between the hexagonal-packing limit and low-density
limit. These interpolations include two fitting constants $\alpha$
and $\gamma$ (see below). The optimum values of these constants
were found by a comparison with MD simulations of a system of
inelastic hard spheres driven by a thermal wall at zero gravity
\cite{Grossman}.

Notice that, prescribing global CRs of \textit{any} type, one
grossly simplifies the delicate issue of phase coexistence that is
expected to occur here in close analogy to the system of
\textit{elastic} hard spheres \cite{Chaikin,Luding}. Still, we
will use the simple CRs \cite{Grossman} to attempt a NSGH
description of the close-packed floating clusters.
In our notation, the CRs \cite{Grossman} read
\begin{equation}
p= n\,T\, \frac{n_c+n}{n_c-n}\,,\,\,\,\,\,\,\kappa=\frac{\mu\,
n\,(\alpha l + d)^2 T^{1/2}}{m^{1/2} \,l}  \label{state}
\end{equation}
and $I=4(\mu/\gamma l)\, q \,n \,m^{-1/2}\, T^{3/2}$. Here $l$ is
the mean free path of the grains,
\begin{equation}
l=\frac{1}{\sqrt{8}nd}\, \frac{n_c-n}{n_c-an}\,, \label{meanfree}
\end{equation}
and $a= 1-(3/8)^{1/2}$. According to Grossman \textit{et al.}
$\alpha=1.15$ and $\gamma = 2.26$. We adopted this value of
$\gamma$, but found better agreement between the hydrodynamics and
MD (see below) for $\alpha = 0.6$. The value of $\mu =
\mathcal{O}(1)$ drops out from the steady-state problem.

Recently a more accurate global EOS $p=p(n,T)$ has been suggested
\cite{Luding}. Still, in the absence of comparably accurate
relations for $\kappa$ and $I$, employing a more accurate EOS in
Eqs. \eqref{govern} would be an excess of accuracy.

Equations \eqref{govern} should be complemented by three boundary
conditions. One of them is $T(0)=T_0=\mbox{const}$. Integrating
the first of Eqs. \eqref{govern} over the height from $0$ to
$\infty$ and using the conservation of the total number of
particles: $\int_0^{\infty}\, n(y)\, dy \, = N/L_x =
\mbox{const}$, we obtain the second boundary condition: $p(0)=m g
N/L_x$. The third one is a zero heat flux (that is, a constant
granular temperature) at $y\to \infty$ \cite{Eggers,tempincrease}.
In practice, one should use the shooting method, varying the heat
flux $- \kappa \, dT/dy$ at $y=0$ until the third condition is
satisfied with desired accuracy.

Let us measure $y$ in units of the gravity length scale
$\lambda=T_0/(m g)$ (note that $\lambda/d$ should be large enough
to fluidize the granulate at the bottom). We rewrite Eqs.
\eqref{govern}, in scaled form, as three first-order equations:
\begin{eqnarray}
\frac{dP}{dy} +\frac{1}{Z}=0\,,\;\;\;\;\frac{d\Phi}{dy} =\Lambda
\,Q(Z) \,P^{3/2}\,,\label{twoeqns}\\ \frac{d}{dy} \left[F_2(Z) \,
P^{3/2}\right]=\frac{\Phi}{F_1(Z)}\,.\label{energy12}
\end{eqnarray}
Here $Z(y)=n_c/n(y)$ is the inverse scaled density,
$P(y)=p(y)/(n_c\,T_0)$ is the scaled pressure, and $-\Phi(y)$ is
the scaled heat flux. The functions $F_1$, $F_2$, and $Q$ are
\begin{eqnarray}
F_1(Z)=\frac{\left[\alpha
Z(Z-1)+\sqrt{32/3}(Z-a)\right]^2}{(Z-a)(Z-1)Z^{2}}\,, \nonumber \\ 
F_2(Z)=\frac{(Z-1)^{3/2}Z^{3/2}}{(Z+1)^{3/2}}\,,\,\,
Q(Z)=\frac{(Z-a) (Z-1)^{1/2}}{(Z+1)^{3/2} Z^{1/2}}\,. \nonumber 
\end{eqnarray}
Finally,
$\Lambda = (64/\gamma) \,q \,(\lambda/d)^2$ and $f=(\sqrt{3}
\,d^2\, N)/(2 \,\lambda\, L_x) $ are two scaled governing
parameters. Parameter $\Lambda$ controls the relative role of the
inelastic heat losses and heat conduction, while $f$ is the
\textit{effective} area fraction of the grains (it can be smaller
or greater than unity). The boundary conditions at the base become
\begin{equation}\label{boundary}
P(0)=f \,\,\,\,\mbox {and}\,\,\,\,
Z(0)=\frac{1+f+(1+6\,f+f^2)^{1/2}}{2\,f}\,.
\end{equation}

\begin{figure}
\centerline{\includegraphics[width=5cm,clip=]{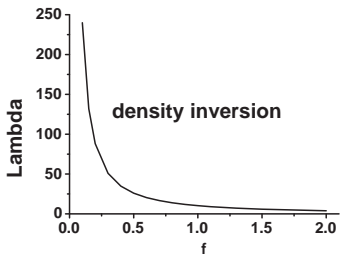}}
\caption{The critical value $\Lambda_c$ of the hydrodynamic
inelasticity parameter $\Lambda$ for a density inversion versus
the effective area fraction $f$. At small $f$ $\Lambda_c$ scales
like $f^{-2}$.} \label{inversion}
\end{figure}

\begin{figure}
\centerline{\includegraphics[width=4cm,bb= 12 16 296 229,
clip=]{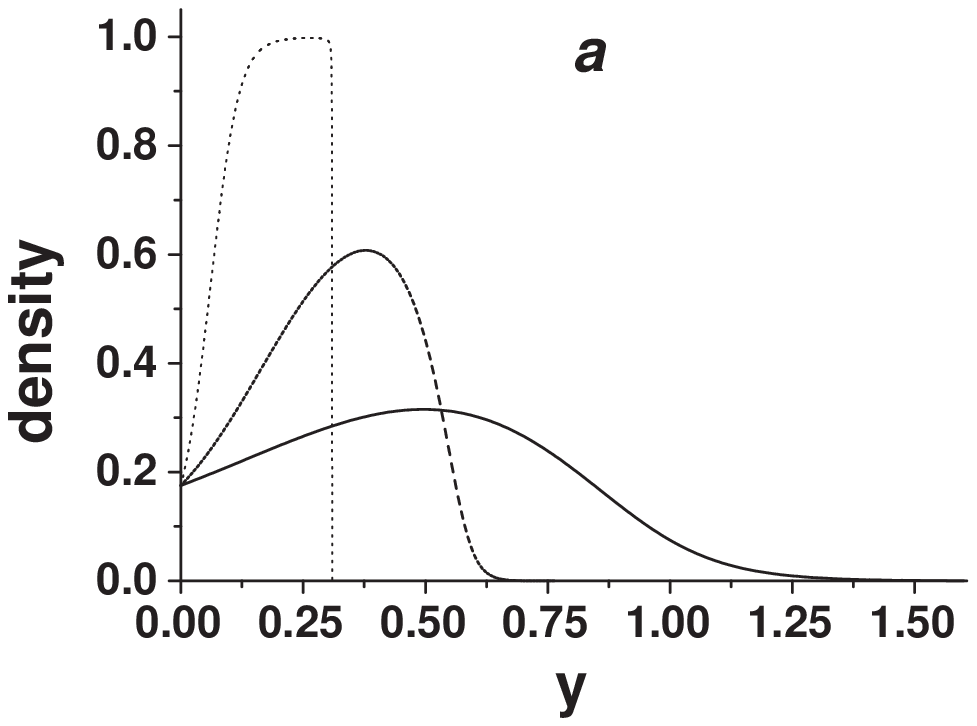}\includegraphics[width=4cm,bb= 12 16
296 229, clip=]{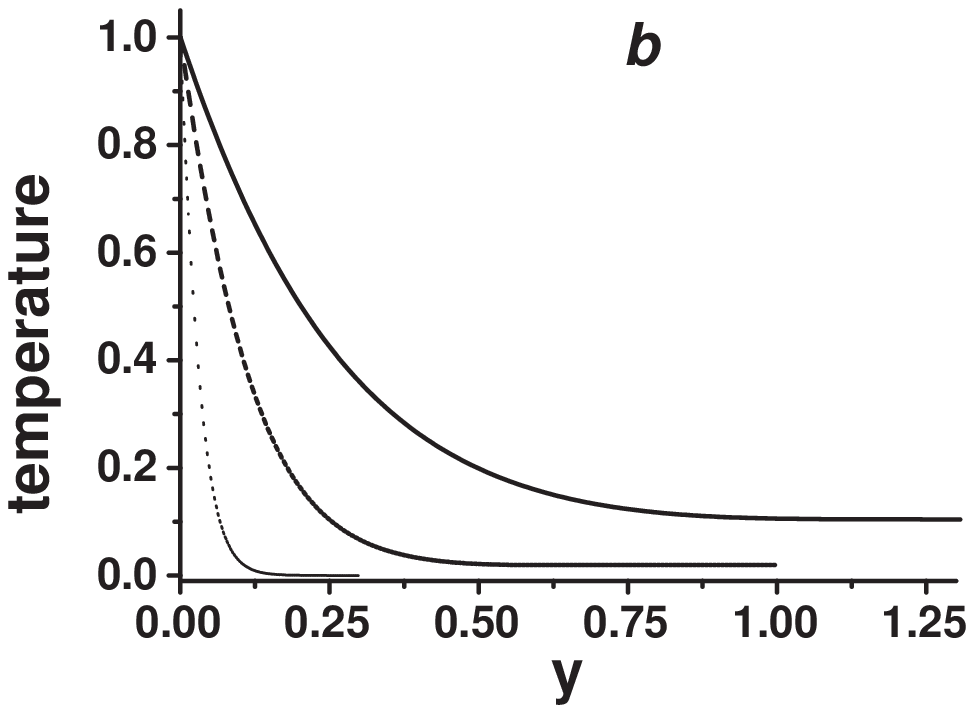}} \caption{Scaled density
(a) and temperature (b) versus the scaled height $y$ for $f=0.25$
and $\Lambda =500$ (solid lines), $2000$ (dashed lines) and
$20,015$ (dotted lines). The dotted lines correspond to the
snapshot shown in Fig. \ref{snapshot}.} \label{threeLambdas}
\end{figure}

Using the hydrodynamic formulation, we first determine the
condition for a density inversion. At too small inelasticity $q$
(the rest of the parameters fixed) there is no density inversion,
like in the elastic case $q=0$ where $T(y)=const$ and $n(y)$ goes
down monotonically. At large enough $q$ the temperature $T(y)$
drops rapidly with $y$. To maintain the hydrostatic balance,
$n(y)$ should \textit{increase} with the height, on an interval of
heights between $y=0$ and the location of the density maximum
$y=y_c$. In our hydrodynamic formulation the density inversion
occurs, at fixed $f$, when $\Lambda>\Lambda_c$, where
$\Lambda_c=\Lambda_c(f)$ is a critical value. The density
inversion is born at $y=0$: $\Lambda=\Lambda_c(f)$ corresponds to
$dn/dy$ vanishing at $y=0$. Using this condition together with
Eqs. \eqref{twoeqns}, \eqref{energy12} and \eqref{boundary}, we
obtain
\begin{equation}\label{Phi0}
 \Phi(0) = -\frac{3}{2} \, f^{1/2}
 \frac{F_1[Z(0)]\,F_2[Z(0)]}{Z(0)}\,.
\end{equation}
For a given $f$, Eq. \eqref{Phi0} prescribes the heat flux at the
base $y=0$ that corresponds to the birth of the density inversion.
Using shooting, we determine, for every $f$, the critical value
$\Lambda_c(f)$, demanding that the temperature approaches a
constant value at large heights. This procedure yields the
critical curve $\Lambda=\Lambda_c(f)$ shown in Fig.
\ref{inversion}.
The density inversion occurs above the critical curve
$\Lambda=\Lambda_c(f)$, and it is more and more pronounced, at
fixed $f$, as $\Lambda$ grows.  Figure \ref{threeLambdas} (a)
shows the density profiles at $f=0.25$ and three different values
of $\Lambda>\Lambda_c$. One can see that, at large enough
$\Lambda$, a hexagonally-packed cluster appears. The scaled
parameters $\Lambda=20,015$ and $f=0.25$ correspond to the
snapshot shown in Fig. \ref{snapshot}. Noticeable is a steep
(exponential) density fall at the upper boundary of the cluster;
the exponent corresponds to the very low temperature there. Figure
\ref{threeLambdas} (b) shows the scaled temperature for these
three cases.

\begin{figure}
\centerline{\includegraphics[width=6cm,clip=]{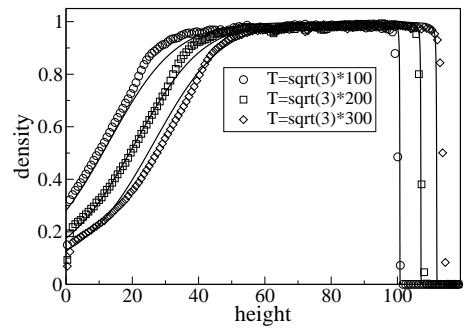}}
\caption{Scaled density versus the height $y$ as predicted by
hydrodynamics (solid curves) and observed in MD simulations, for
three different values of the temperature at the base. The
parameters are described in the text. The height is measured here
in units of $d$.} \label{3profiles}
\end{figure}

Figure \ref{3profiles} compares the density profiles, predicted by
this hydrodynamics (solid curves), with the profiles found in MD
simulations with $N=10^4$ particles of diameter $d=1$, mass $m=1$
and $r=0.98815$. The (periodic) box width is $L_x=100$, the
gravity acceleration $g=1$. The MD simulations were done for three
different values of the temperature at the base: $T_0=100
\sqrt{3},\, 200 \sqrt{3}$ and $300 \sqrt{3}$. The hydrodynamic
parameters in these three cases are $f=0.5$ and $\Lambda=5004$;
$f=0.25$ and $\Lambda=20015$, and $f=0.167$ and $\Lambda=45036$,
respectively. One can see that the agreement between hydrodynamics
and MD simulations is surprisingly good.

\begin{figure}
\centerline{\includegraphics[width=5.0 cm,clip=]{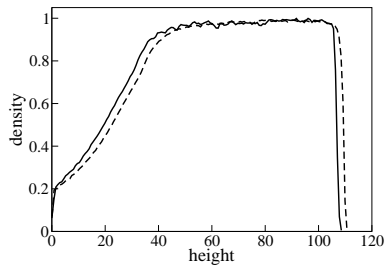}}
\caption{Checking the hydrodynamic scaling. The solid line
corresponds to the MD simulation shown in Fig. 1. The dashed line
corresponds to another simulation: with $N=2 \cdot 10^4,\,
L_x=100,\, T_0=4\sqrt{3}\cdot 10^2,\, r=0.997051$ and $g=d=m=1$.
For the dashed line the height $y$ is shrunk by a factor of 2.}
\label{rescaling}
\end{figure}

An additional argument in favor of hydrodynamics follows from the
dimensional analysis of the problem. The full set of parameters
includes $d,m,r,g,T_0,N$ and $L_x$. One can always choose
$d=m=g=1$, so there are actually \textit{four} independent
parameters. This number reduces to \textit{three} for an
$x$-independent steady state, as $N$ and $L_x$ enter the problem
only through $N/L_x$. It is crucial that hydrodynamics further
reduces the number of parameters: now only \textit{two} scaled
parameters $\Lambda$ and $f$ appear. This prediction is very
robust, as it is independent of the particular form
of the functions $F_1,\,F_2$ and $Q$ (and of the values of
$\alpha$ and $\gamma$). We verified this prediction in MD
simulations by varying $N, T_0$ and $r$, but keeping
$\Lambda=20,015$ and $f=0.25$ constant. After rescaling the
coordinate $y$ by $\lambda=T_0/(m g)$, the resulting density
profiles almost coincide with each other, see Fig.
\ref{rescaling}. A small shift between the two profiles, observed
in Fig. \ref{rescaling}, is apparently caused by small vertical
oscillations of the granulate. In this example the oscillation
amplitude is about 3 particle diameters \cite{oscillations}.

As already mentioned, the global CRs completely ignore the issue
of coexistence, beyond the freezing point, of different phases of
the granulate: the liquid-like phase, the random close-packed
phase \textit{etc}. So why are they so successful? We believe the
reason is the following. The vibrofluidized steady state,
considered in this work, has a zero mean flow. Therefore, the
viscosity  terms in the hydrodynamic equations vanish. This fact
is not merely a technical simplification. The shear viscosity of
granular flow is finite in the liquid-like phase, and
\textit{infinite} in the (multiple) domains of the random
close-packed phase. The effective \textit{total} viscosity of the
system is expected to \textit{diverge} when the coarse-grained
density slightly exceeds the freezing density. This invalidates
\textit{any} NSGH for sufficiently dense flows, and necessitates
the introduction of an order parameter and a different type of the
stress-strain relation into the theory, \textit{cf.} Ref.
\cite{Aranson1}. Luckily, these complications do not appear for a
zero-mean-flow state. Indeed, the EOS, heat conductivity and
inelastic heat loss rate do not exhibit any singularity around the
freezing point, and all the way to the hexagonal close packing.
Therefore, the NSGH remains reasonably accurate far beyond the
freezing point. A future work should address the important
question about the range of applicability of the NSGH (actually,
of any binary collision model) for solid yet vibrated phase,
versus the granular statics approach.

We are grateful to Stefan Luding for comments and for sharing with
us his unpublished results on the shear viscosity divergence. We
thank Igor Aronson, Detlef Lohse and Arkady Vilenkin for
discussions. This research was supported by the Israel Science
Foundation (Grant No. 180/02) and by Deutsche
Forschungsgemeinschaft (Grant PO 472/6-1).


\begin{references}
\bibitem{defects} One can also clearly see non-hydrodynamic structures: domains of different crystalline orientation, domain
boundaries and point defects, analogous to those observed in
polycrystals. These are also observed when inelastic hard spheres
are driven by a thermal wall at zero gravity \cite{Poeschel}.
\end{references}
\end{document}